\documentclass[10pt]{article}

\usepackage{amsmath}
\usepackage{amssymb}

\usepackage{graphicx}

\usepackage{cite}

\usepackage[labelfont=bf,labelsep=period,justification=raggedright]{caption}

\usepackage {textcomp}
\usepackage{color}
\usepackage{enumerate}
\usepackage{amsmath}
\usepackage{amssymb}
\usepackage[mathscr]{eucal}
\usepackage{bm}
\usepackage{subfigure}
\usepackage{theorem}
\usepackage{graphicx}
\usepackage{setspace}
\usepackage{amssymb,amsfonts,amsmath}

\usepackage{xr}

\usepackage[T1]{fontenc}
\usepackage[utf8]{inputenc}
\usepackage{authblk}

\usepackage{lipsum}

\makeatletter
\renewcommand{\@biblabel}[1]{\quad#1.}
\makeatother

\date{}

\begin{document}

\begin{flushleft}
{\Large
\textbf{Unraveling Adaptation in Eukaryotic Pathways: Lessons from Protocells}
}

%


Giovanna De Palo$^{1,2}$, 
Robert G. Endres$^{1,2,\ast}$, 
\\
\bf{1} Department of Life Sciences, Imperial College, London, United Kingdom
\\
\bf{2} Centre for Integrative Systems Biology and Bioinformatics, Imperial College, London, United Kingdom
\\
$\ast$ E-mail: r.endres@imperial.ac.uk
\end{flushleft}

%

\section*{Abstract}
Eukaryotic adaptation pathways operate within wide-ranging environmental conditions without stimulus saturation. Despite numerous differences in the adaptation mechanisms employed by bacteria and eukaryotes, all require energy consumption. Here, we present two minimal models showing that expenditure of energy by the cell is not essential for adaptation. Both models share important features with large eukaryotic cells: they employ small diffusible molecules and involve receptor subunits resembling highly conserved G-protein cascades.  Analyzing the drawbacks of these models helps us understand the benefits of energy consumption, in terms of adjustability of response and adaptation times as well as separation of cell-external sensing and cell-internal signaling. Our work thus sheds new light on the evolution of adaptation mechanisms in complex systems.

\section*{Author Summary}
Adaptation is a common feature in sensory systems, well familiar to us from light and dark adaptation of our visual system. Biological cells, ranging from bacteria to complex eukaryotes, including single-cell organisms and human sensory receptors, adopt different strategies to fulfill this property. However, all of them require substantial amounts of energy to adapt. Here, we compare the different biological strategies and design two minimal models which allow adaptation without requiring energy consumption. Schemes similar to the ones we proposed in our minimal models could have been adopted by ancient protocells, that have evolved into the pathways we now know and study. Analyzing our models can thus help elucidate the advantages brought to the cells by consumption of energy, including the bypassing of hard-wired cell parameters such as diffusion constants with increased control over time scales.

\section*{Introduction}

The ability to adapt to different environmental conditions is a hallmark of any living organism, allowing sensory systems to adjust their sensitivity to changes in nutrient availability, stress, and other stimuli \cite{Fain_2003_book,Barkai_1997}.
Adaptation mechanisms generally rely on biochemical feedback pathways, specific for the organism and type of stimulus \cite{Yi_2000,Alon_1999}.
Surprisingly, when comparing adaptation pathways in different organisms, one cannot help noticing the wide variety of pathway designs, ranging from remarkably simple in bacterial chemotaxis \cite{Sourjik_2010} to highly complex in eukaryotes, including {\it Dictyostelium discoideum} chemotaxis \cite{Manahan_2004}, olfactory-transduction \cite{Kaupp_2010, Kleene_2008} and photo-transduction \cite{Yau_2009}.
Since bacterial chemotaxis  is known to work (and adapt) exquisitely well, it is unclear why eukaryotic organisms have such elaborated pathway designs, and what their advantages might be.

In light of such complexity, the ``physicist's approach'' may help guide the identification of common principles among the different pathways. 
Recently, Lan {\it et al.} \cite{Lan_2012} analyzed a core adaptation pathway and concluded that adaptation always relies on energy consumption by the cell.
However, whether it is generally impossible to adapt without consuming energy is still an open question, considering how adaptation mechanisms could have evolved in ancient protocells without sophisticated pathways.
We address this issue by first comparing the sensory pathways of small bacterial and large eukaryotic cells, highlighting their similarities and differences. We then investigate the possibility of achieving adaptation with no energy consumption by means of a ``protocell'' based on equilibrium physics.
The hypothesis of no cost of energy for the cell guides us in designing minimal adaptation mechanisms. 
By pondering the drawbacks of these mechanisms and the advantages brought by energy-consuming pathways, we unravel some of the complexity of sensory systems.
These considerations may provide a path towards a general theory of adaptation in eukaryotic cells.

\section*{Results}

\subsection*{Local control in bacteria, global control in eukaryotes}

\begin{figure}[htp]
 \begin{center}
\includegraphics{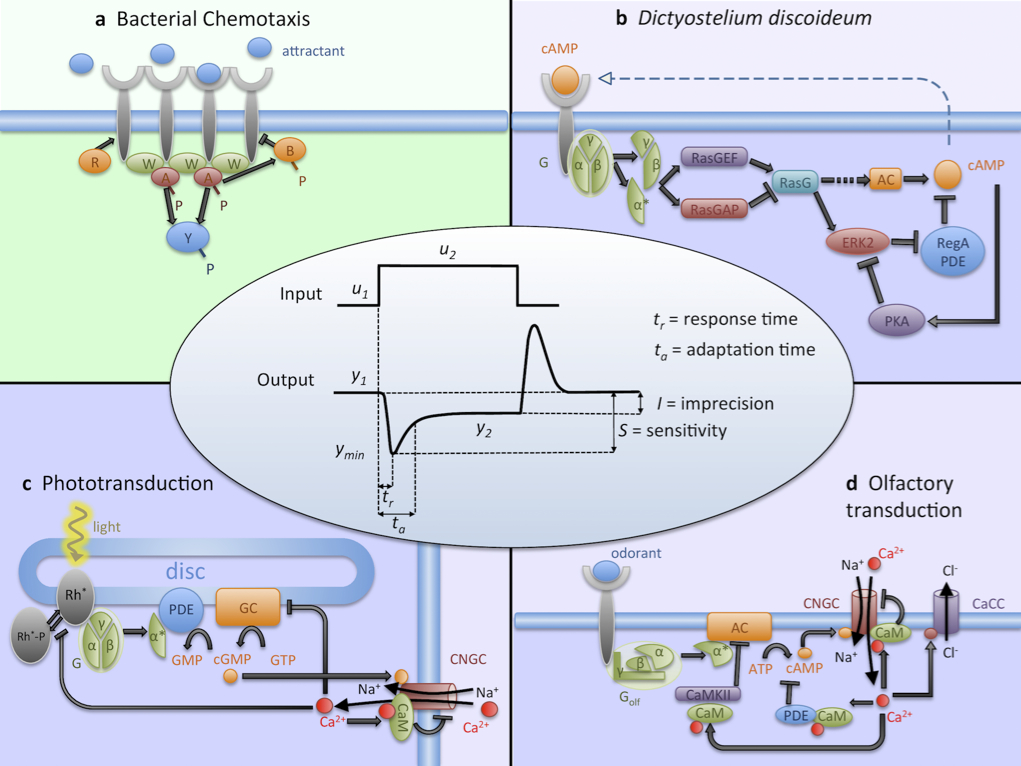}
\caption{{\bf Schematic description of adaptation pathways and their properties.} (a) Bacterial chemotaxis. Attractant molecules bind to chemoreceptors, which cluster due to adapter protein CheW and kinase CheA, responsible for phosphorylation of CheY and hence regulation of the flagellar motors. CheR and CheB (activated by CheA) mediate adaptation by methylating and demethylating receptors, respectively \cite{Kentner_2009}. (b) {\it Dictyostelium} chemotaxis, as related to the production and sensing of cAMP. After the binding of cAMP to the receptor, the G-protein complex dissociates \cite{Manahan_2004} and RasG protein is activated \cite{Takeda_2012, Iglesias_2012b}. The adenylyl cyclase (AC), possibly through the ${\it PI(3,4,5)P_3}$ pathway \cite{Manahan_2004, Kimmel_2003}, is activated, and produces cAMP, secreted for cell aggregation. The concentration of cAMP is also increased through ERK2, which inhibits the phosphodiesterase (PDE) RegA, in turn hydrolyzing cAMP. This pathway is inhibited by PKA, activated by cAMP \cite{McMains_2008, Valeyev_2009, Kimmel_2003}. (c) Photo-transduction. Light activates rhodopsin (${\it Rh^*}$) and, following a G-protein cascade, phosphodiesterase hydrolyzes cGMP. At low PDE levels (in the dark), cGMP allows the influx of ${\rm Ca^{2+}}$ through cyclic nucleotide-gated channels (CNGC). Adaptation is mediated via ${\rm Ca^{2+}}$-dependent feedback loops: inhibition of GC, phosphorylation of ${\it Rh^*}$, and CNGC \cite{Yau_2009}. (d) Olfactory-transduction. Odorant binds to the G-protein coupled olfactory receptor, activating AC to produce cAMP, causing the opening of CNGC. The influx of ${\rm Ca^{2+}}$ opens chloride channels to amplify signal. Several ${\rm Ca^{2+}}$-dependent feedback mechanisms mediate adaptation: inhibition of AC, CNGC and cAMP \cite{Kleene_2008, Kaupp_2010}. (central panel) Response of an adaptive system to a step stimulus, and characteristic features considered here, with $I=\left|((y_2-y_1)/y_1)\right|$ and $S=\left|((y_{min}-y_1)/y_1)/((u_2-u_1)/u_1)\right|$ \cite{Ma_2009}.
}
 \label{fig:pathways}
 \end{center}
 \end{figure}

Constrained by their small size, bacteria are spatially highly organized \cite{Sourjik_2010,Shapiro_2009}.
Consider e.g. bacterial chemotaxis in {\it Escherichia coli}, known for amplifying weak signals by cooperative receptors and precise adaptation (Fig.~\ref{fig:pathways}a). There is an apparent high level of local control in these processes: receptors are arranged hexagonally in clusters \cite{Briegel_2012} and the adaptation enzymes CheR and CheB are tethered to the receptors to increase local enzyme concentration and specificity, and to reduce noise \cite{Endres_2006_assistance_neighborhood}.
Adaptation in bacterial chemotaxis is a robust feature of the pathway, without ``fine-tuning'' of biochemical parameters \cite{Barkai_1997}, achieved by integral feedback control \cite{Yi_2000}.
Integral feedback control \cite{Lan_2012, DePalo_2013} and the related incoherent feedforward loop \cite{Takeda_2012} are also found in eukaryotes.

Despite these similarities, eukaryote's most striking feature is the staggering complexity, often relying on long, multistep signaling cascades in parallel \cite{Cai_2011}. Consider three different pathways: chemotaxis in {\it Dictyostelium discoideum} (Fig.~\ref{fig:pathways}b), and photo- (Fig.~\ref{fig:pathways}c) and olfactory-transduction (Fig.~\ref{fig:pathways}d) in mammalian sensory systems. 
Perhaps most perplexing, the key signaling component for all of them are small, fast-diffusible second messengers, like ${\rm Ca^{2+}}$ for olfactory- and photo-transduction (which is even toxic for the cell in large quantities \cite{Dodd_2010,Mattson_2003}), and cAMP in {\it Dictyostelium} amoeba. Relying on these small signaling molecules could negatively affect the precision of the response. Due to their fast diffusion, affected targets may not only be the designated molecules, but large parts of the cells.
Moreover, all of the eukaryotic examples mentioned rely on G-protein coupled receptors (GPCR).
The excitation of the receptor catalyzes the production of GTP and the dissociation of the G-protein. GTP then binds to the $\alpha$ subunit and the $\beta\gamma$ subunits (for {\it Dictyostelium} \cite{McMains_2008,Manahan_2004}) or the $\alpha$ subunit (for photo- and olfactory-transduction \cite{Kaupp_2010,Yau_2009}) can activate the downstream processes.
The activity of the guanosine triphosphatase (GTPase) hydrolizes the GTP which detaches from the $\alpha$ subunit. Subsequently, the $\alpha$ subunit can re-bind the $\beta\gamma$ subunits, thus reassembling the G-protein \cite{Janetopoulos_2001}.

All bacterial and eukaryotic adaptation pathways share the consumption of energy by hydrolysis of fuel molecules, including S-adenosyl methionine (SAM) in bacterial chemotaxis, and cyclic adenosine and guanosine monophosphate (cAMP and cGMP) in eukaryotes \cite{Lan_2012}. These observations led to the conclusion that energy consumption is an essential ingredient in precise adaptation \cite{Lan_2012} (see Fig.~\ref{fig:pathways}, middle panel, for a definition).
However, ancient protocells might have been able to respond and adapt to stimuli without this requirement, with molecular components added later by evolution, to produce the currently observed pathways.
As it turns out, our protocell models, which only rely on equilibrium physics for the cellular components, may help unravel some of the signaling complexity in eukaryotic cells.

\subsection*{Equilibrium adaptation models of protocells}

\begin{figure}[htp]
 \begin{center}
\includegraphics{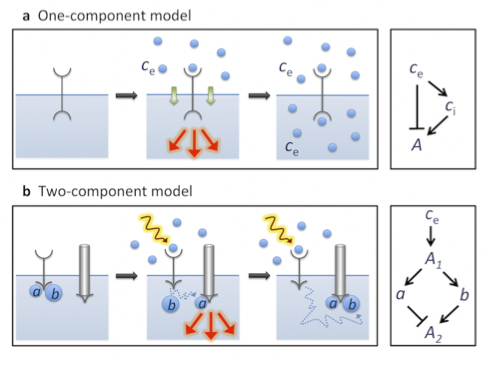}
\caption{{\bf Minimal adaptation models.} (a) One-component model. (left) The receptor on the cell membrane has two ligand-binding sites, one extra- and one intracellular. Upon extracellular ligand binding, the receptor signals, but stops when ligand has permeated the membrane and binds intracellularly to the receptor. (right) Schematic diagram of corresponding incoherent feedforward loop with $c_i$ the slow intermediate species mediating adaptation. (b) Two-component model. (left) The first component is a receptor, comprising an extracellular binding site for ligand, and an intracellular domain to which two subunits, $a$ and $b$, are bound. The second component is responsible for downstream signaling. When the receptor is stimulated (e.g. by ligand or light) the $a$ and $b$ subunits are released and diffuse towards the second component. Since diffusion of $a$ is faster than that of $b$, $a$ binds first to the second component, which starts signaling until $b$ binds. (right) Schematic diagram of corresponding incoherent feedforward loop with two intermediate species $a$ and $b$. Species $b$ diffuses more slowly than $a$ and mediates adaptation.
}
 \label{fig:models}
 \end{center}
 \end{figure}

Here, we present two minimal models in which a protocell exploits the non-equilibrium aspect of the changing external environment to respond and adapt to a stimulus without consuming (dissipating) energy itself.
The first model contains only one component, represented by a receptor on the cellular membrane (Fig.~\ref{fig:models}a, left). This receptor includes two sensing regions: the first binds extracellular ligand, the second mediates intracellular sensing and adaptation. As soon as the (extracellular) ligand arrives at the cell surface, the receptor binds and starts signaling. However, the stimulus molecules are also able to permeate the cellular membrane, e.g. via passive pores, and to consequently bind additionally to the intracellular region of the receptor. This second binding blocks the signaling activity of the receptor, thus precisely counteracting the activation due to the extracellular binding.

In case the external stimulus cannot be used to mediate adaptation, e.g. for photo-transduction, a second model with two components is needed (Fig.~\ref{fig:models}b, left). The first represents the receptor, which senses the extracellular stimulus and, upon stimulation, releases two intracellular subunits, $a$ and $b$. The $a$ subunit is smaller and can diffuse faster than the $b$ subunit. The second component is a membrane-bound protein, able to respond to the binding of the $a$ and $b$ subunits. In particular, the binding of $a$ to this second protein causes an increase of its signaling activity, while the binding of $b$ exactly compensates for it, and hence turns signaling off.

\subsection*{Precise adaptation at no cost}

\begin{figure}[htp]
 \begin{center}
\includegraphics{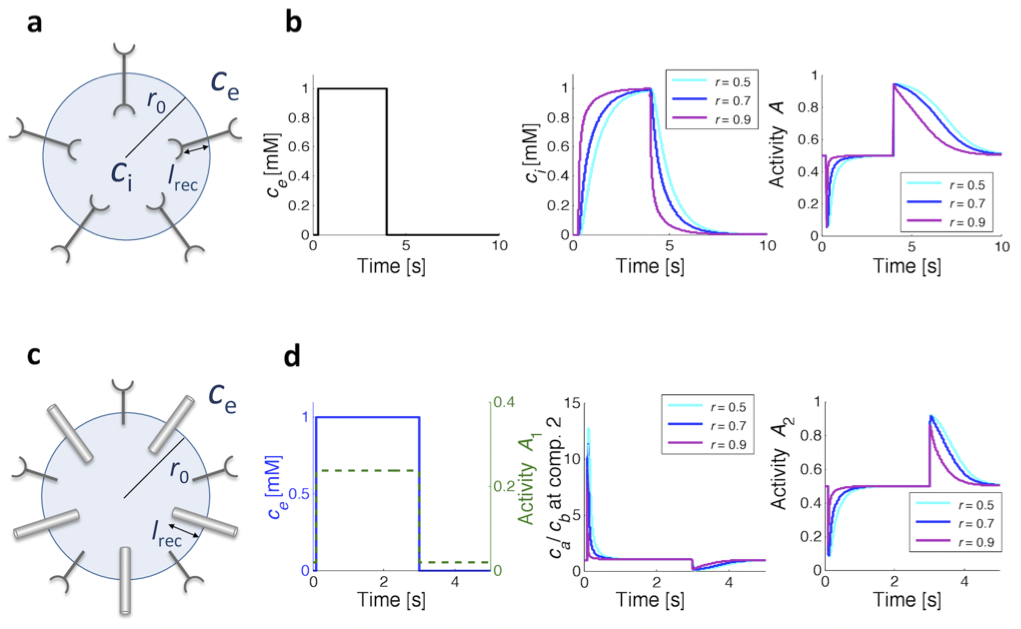}
\caption{{\bf Implementation of the minimal models in protocells.} (a) One-component model, with $c_e$ and $c_i$ the extracellular and intracellular concentrations, $r_0$ the radius of the cell, and $l_{rec}$ the length of the receptor. (b) Simulation results for the one-component model showing time courses of stimulus $c_e$ (left), inner concentrations $c_i$ (middle), and activity (right) for different relative receptor lengths $r=(r_0-l_{rec})/r_0$, with $r_0 = 5$ $\mu$m, $D= 3$ $\mu$m$^2$/s, $K_{1}=0.02$ mM and $K_{2}= 0.5$ mM. (c) Two-component model, with $l_{rec}$ the length of the second component. (d) Simulation results for the two-component model. Shown are outer concentration $c_e$ (left axis) and activity of the first receptor (right axis) (left), $a$ and $b$ concentration ratio as bound to the second component (middle), and activity of the second component (right) for different relative receptor lengths $r$, with $r_0 = 5$ $\mu$m, $D_{a}= 300$ $\mu$m$^2$/s and $D_{b} = 8$ $\mu$m$^2$/s, $K_D^{\rm on} = 0.02$ mM, $K_D^{\rm off} = 0.5$ mM, and $L_{D} = 0.02$ mM.
}
 \label{fig:main_results}
 \end{center}
 \end{figure}

{\bf One-component model.} Let us assume the cell is initially equilibrated to concentration $c_0$ and then the external concentration is suddenly changed to $c_e$. The resulting difference in concentration between inside and outside the cell equilibrates following the solution of the diffusion equation with spherical symmetry (hypothesizing that the cellular membrane is sufficiently permeable)\cite{Dutta_book}
\begin{eqnarray}
&&\frac{c_i(t)-c_e}{c_0-c_e}=-\frac{2}{\pi r}\sum_{m=1}^\infty \frac{(-1)^m}{m}\cdot \sin(m\pi r)e^{\frac{-m^2\pi^2Dt}{{r_0}^2}},
\label{eq:diffusion}
\end{eqnarray}
where $c_i(t)$ is the internal concentration at time $t$, $c_0$ and $c_e$ are the respective initial internal and external concentrations, $r_0$ is the radius of the cell, $D$ the diffusion coefficient, and $r=(r_0-l_{rec})/r_0$ the normalized radius corresponding to the inner length of the receptor $l_{rec}$ (see Fig.~\ref{fig:main_results}a for a schematic).
We further assume that external ligand binding favors the receptor \textit{off} (inactive) state (with ligand dissociation constants ${K_{e}}^{\rm off}\ll{K_{e}}^{\rm on}$) and that the internal ligand binding favors the \textit{on} (active) state (${K_{i}}^{\rm on}\ll{K_{i}}^{\rm off}$), in order to compensate for external ligand binding during adaptation.
For simplicity, we consider ${K_{e}}^{\rm off}={K_{i}}^{\rm on}=K_{1}$ 
and ${K_{e}}^{\rm on}={K_{i}}^{\rm off}=K_{2}$.
The corresponding free-energy difference of the single receptor and the activity associated with it become (using equilibrium Boltzmann statistics) \cite{Endres_2006_assistance_neighborhood}
\begin{eqnarray}
&\Delta f=&f_{\rm on}-f_{\rm off}=\ln\left(\frac{1+c_e(t)/K_{1}}{1+c_e(t)/K_{2}}\right)+\ln\left(\frac{1+c_i(t)/K_{2}}{1+c_i(t)/K_{1}}\right)
\label{eq:df_mod1}
\\
&A=&\frac{1}{1+e^{\Delta f}}
\label{eq:A_mod1}
\end{eqnarray}
with energy in units of $k_BT$.

Figure~\ref{fig:main_results}b shows adaptation of the one-component model to an extracellular concentration-step change as input, $c_e(t)$ (left). The intracellular concentration response $c_i(t)$ due to the diffusive process for different lengths of the inner side of the receptor (middle) allows the activity output, $A$, to adapt precisely at no cost for the cell (see Figs.~S1-S4 for additional results).

{\bf Two-component model.} The second model also relies on equilibration for adaptation, albeit by an all-internal mechanism (see Fig.~\ref{fig:main_results}c). The external concentration $c_e$ becomes the input for the first component - the receptor, which responds to the transmembrane free-energy difference by changing its activity. In particular, its free-energy difference and activity are, respectively,
\begin{eqnarray}
&&{\Delta f_{1}}=\sigma + \ln \left(\frac{1+c_e(t)/{K_D}^{\rm off}}{1+c_e(t)/{K_D}^{\rm on}}\right)
\\
&&A_{1}=\frac{1}{1+e^{\Delta f_{1}}}.
\label{eq:act2_main}
\end{eqnarray}
Parameter $\sigma >0$ represents the bias towards the off state in absence of external stimulus, and the contribution of the binding energy of the $a$ and $b$ subunits on the cytoplasmic domain of the receptor.
The concentration of $a$ and $b$, released by the first component, is set proportional to the activity $A_{1}$.

The solution of the diffusion equation, Eq.~\eqref{eq:diffusion}, is then used to represent the diffusive process of $a$ and $b$ molecules from the receptor to the second component. The free-energy difference and the activity of the second component are calculated as
\begin{eqnarray}
&&{\Delta f_{2}}=\ln \left(\frac{1+\frac{c_a(t)}{{L_D}}}{1+\frac{c_b(t)}{{L_D}}}\right)
\label{eq:df2_model2}\\
&&A_{2}=\frac{1}{1+e^{\Delta f_{2}}},
\label{eq:A2_model2}
\end{eqnarray}
where, for simplicity, we assume $K_{Da}^{\rm off}=K_{Db}^{\rm on}=L_{D}$ and $K_{Da}^{\rm on}=K_{Db}^{\rm off}=\infty$ (see Text S1 for details).

Figure~\ref{fig:main_results}d shows adaptation of the two-component model. The outer concentration is followed by activity $A_{1}$ of the first component, which does not adapt (left). However, the ratio in the concentrations of $a$ and $b$, as sensed by the second component after diffusion (middle), allows the activity $A_{2}$ to adapt precisely (right).
Hence, similar to the one-component model, also the two-component model achieves precise adaptation, just relying on diffusion of the $a$ and $b$ subunits at no energy cost (see Figs.~S8-S11 for additional results).

\begin{figure}[htp]
 \begin{center}
\includegraphics{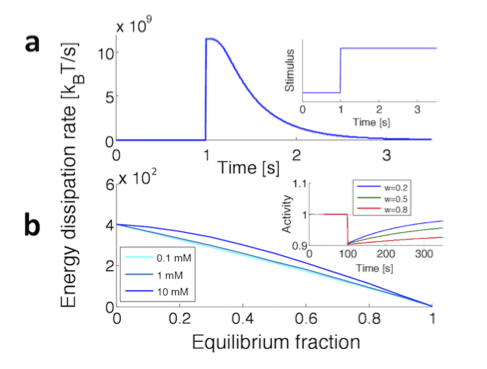}
\caption{{\bf Comparison of energy-dissipation rates.} (a) One-component model in response to the step stimulus depicted in inset, with a concentration-step change from 1 to 1.5 mM. (b) Steady-state value for the bacterial chemotaxis (BC) model described in {\it Methods} and corresponding activity profiles (inset). Equilibrium fraction (parameter $w$  in inset) represents ratio of equilibrium and non-equilibrium contributions (see {\it Methods}).
}
 \label{fig:entropy}
 \end{center}
 \end{figure}

{\bf Energy dissipation.} Our models do not require energy consumption, but achieve adaptation by means of the {\it external} energy provided by the stimulus. Figure~\ref{fig:entropy}a shows the resulting energy-dissipation rate of the one-component model, which confirms that most energy is dissipated right after the step change with the rate decreasing to zero during equilibration. In contrast, a biologically regulated system constantly dissipates internal energy. To illustrate this, we consider the linear bacterial chemotaxis model introduced in \cite{Clausznitzer_2010}
\begin{equation}
\frac{dm}{dt}=g_R(1-A)-g_B A
\label{eq:chem_model}
\end{equation}
with $m$ the receptor-methylation level, $g_R$ and $g_B$ the methylation and demethylation rate constants depending on enzymes CheR and CheB, and $A$ the receptor activity.
To obtain a precisely adapting non-equilibrium process, we add reverse reactions to Eq.~\eqref{eq:chem_model}.
To compare this with an imprecisely adapting equilibrium process we use a similar equation, which, however, now depends on $m$ instead of $A$ (see {\it Methods} for a detailed description of this model).
From Fig.~\ref{fig:entropy}b it emerges clearly that in bacterial chemotaxis there is a trade-off between energy consumption and precision in adaptation, i.e. energy consumption is required to achieve a high degree of precision \cite{Lan_2012}.

\begin{figure}[htp]
 \begin{center}
\includegraphics{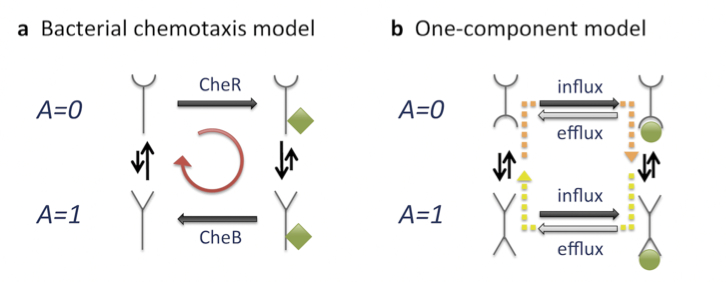}
\caption{{\bf Equilibrium vs. non-equilibrium processes in adaptation.} (a) Bacterial chemotaxis model with two activity states ($A=0$ and 1). For simplicity only one methyl-group (green diamond) is shown, added by CheR and removed by CheB. The red arrow represent the futile cycle undertaken when adapted. (b) One-component model with different receptor states. {\it Influx} and {\it efflux} describe the continuous entering and leaving of ligand, and the green disc represents internally bound ligand (external ligand is not shown). When adapted all individual reactions are equilibrated and satisfy detailed balance. Dashed arrows represent the breaking of detailed balance in response to an increase (orange arrow) or a decrease (yellow arrow) of external ligand concentration.
}
 \label{fig:entropy_cycle}
 \end{center}
\end{figure}

Figure~\ref{fig:entropy_cycle} illustrates the sources of energy dissipation.
In bacterial chemotaxis (Fig.~\ref{fig:entropy_cycle}a), receptors are constantly being modified, even in the adapted steady state.
At a given ligand concentration, demethylated receptors tend to be inactive. Once inactive, receptors become methylated by the action of CheR. Once methylated, receptors tend to be active, and subsequently become demethylated by CheB. The whole process represents a futile cycle, driven by the (nearly) irreversible methylation and demethylation reactions, which do not satisfy detailed balance.
In stark contrast, in our one-component model depicted in Fig.~\ref{fig:entropy_cycle}b, all reactions are equilibrated when adapted (and similarly in our two-component model).
Specifically, inactive receptors are ``modified'' by the influx of ligand, which after binding to the receptors establishes an equilibrium between active and inactive states. The influx is, however, counterbalanced by the efflux (and thus the unbinding) of ligand, which again is accompanied by equilibration of the receptors.
Only when the external ligand concentration changes, detailed balance is temporally broken.
When the external concentration of ligand is increased, the activity drops, followed by an influx of the ligand through the membrane. This is followed by ligand binding on the intracellular site of the receptor, which then restores the equilibrium activity state (orange dashed line).
If the extracellular concentration is subsequently decreased back to the initial level, the activity increases, followed by an efflux of the ligand and restoration of the equilibrium activity state (yellow dashed line).
Any energy dissipation  is paid for by the environment, not by the cell.

\subsection*{Adaptation time and fold-change detection}

How do these minimal models compare with data from actual adaptation mechanisms?
The {\it adaptation time}, a measure of the speed of adaptation, can be defined as the time required for the response to return back to half of the displacement from the prestimulus value (see Fig.~\ref{fig:pathways} for a graphical explanation).
Experimentally measured adaptation times vary from seconds to minutes: adaptation in bacterial chemotaxis by receptor methylation can take up to hundreds of seconds for very large stimuli \cite{Berg_1975, Keymer_2006} and similarly for cell-internal adenylyl cyclase ACA in {\it Dictyostelium} chemotaxis \cite{Brzostowski_2013} (although cGMP and activated RasG can be significantly faster \cite{Valkema_1994,Takeda_2012}).
In contrast, adaptation of the transmembrane currents in the olfactory- \cite{Menini_1995} and photo-transduction \cite{Torre_1995} pathways is faster (a few seconds).

Another important feature of adapting systems is {\it fold-change detection} (FCD), which allows cells to interpret chemical gradients irrespective of scale \cite{Shoval_2010}. Specifically, when applied to a step change in concentration, the output response should only depend on the fold change in the input; if the input is rescaled by a multiplicative factor, the output should remain exactly the same for every time point considered. This feature entails both exact adaptation (that the system returns exactly to the prestimulus value) and Weber's law (that the smallest detectable stimulus is proportional to the background stimulus), but it is not implied by either or both of them. Bacterial chemotaxis, despite considerable energy consumption, indeed exhibits fold-change detection \cite{Lazova_2011}. Olfactory- and photo-transduction do not adapt perfectly, and thus do not satisfy FCD.

\begin{figure}[htp]
 \begin{center}
\includegraphics{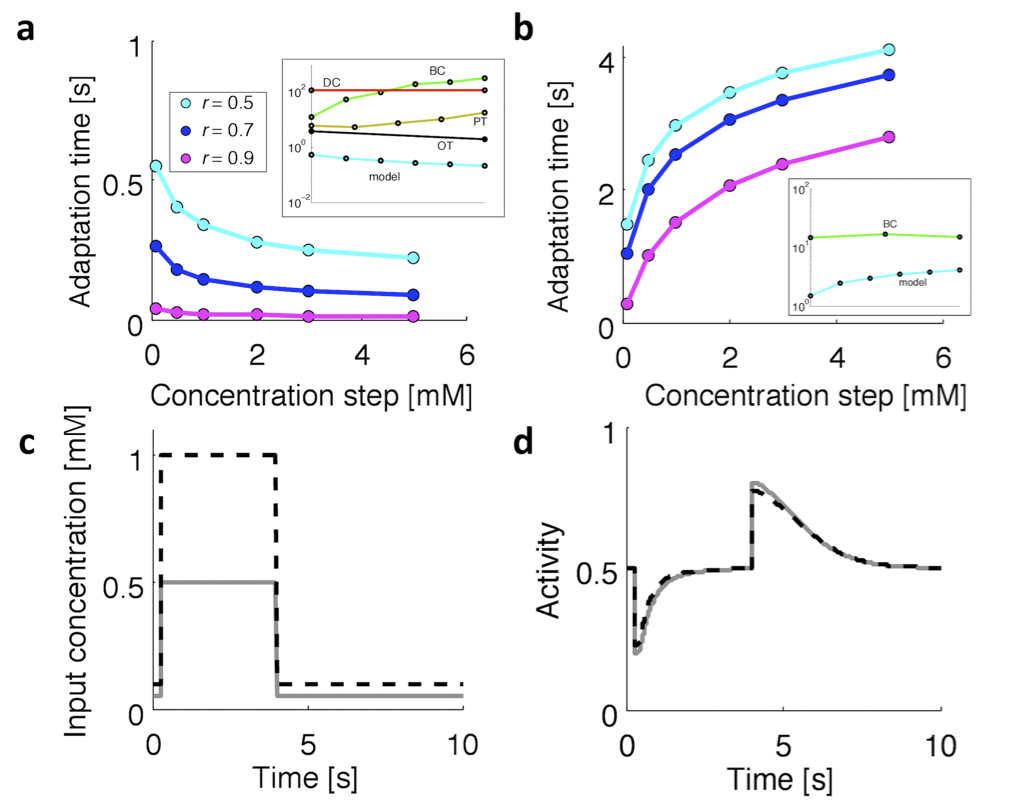}
\caption{{\bf Adaptation time and fold-change detection.} (a) Adaptation time of the one-component model in response to a positive step of increasing size for different relative receptor lengths (from 0 mM to 0.1, 0.5, 1, 2, 3, 5 mM). (inset) Experimental adaptation-time data from bacterial chemotaxis (BC, tumble bias from Fig.~2 of \cite{Min_2012}), {\it Dictyostelium} (DC, adenylyl cyclase ACA from Fig.~4C of \cite{Brzostowski_2013}), photo-transduction (PT, transmembrane current from Fig.~6B of \cite{Forti_1989}), olfactory-transduction (OT, transmembrane current from Fig.~1a of \cite{Menini_1995}). (b) Adaptation time of the one-component model in response to a negative step of increasing step size (same concentrations as in (a)). (inset) Comparison with the experimental BC data.
(c-d) Input concentration (c) and activity (d) for a step from 0.1 to 0.5 mM (gray line) and a step from 0.2 to 1 mM (black dashed line). Although the input concentration is doubled, the two activity profiles superimpose almost exactly.
}
 \label{fig:times_FCD}
 \end{center}
 \end{figure}

Figure~\ref{fig:times_FCD} shows the results for the adaptation time and FCD for the one-component system (see also Figs.~S5-S7; the two-component behaves very similarly, see Figs.~S12, S13). The adaptation time in response to a positive step decreases with increasing size (Fig.~\ref{fig:times_FCD}a), while the response to a negative step has the opposite behavior (Fig.~\ref{fig:times_FCD}b).
This can be explained within our intuitive, minimal model: since $K_e^{\it off}\ll K_e^{\it on}$ for extracellular binding (and vice versa for the intracellular), in response to a positive step, ligand strongly binds to the off-state and weakly to the on-state of the extracellular domain of the receptor. As a result, the state of the receptor switches from {\it on} to {\it off}. When the ligand enters the cell, even a small concentration is enough to bind intracellularly the receptor in the on-state, turning it on. Therefore adaptation is fast, and the adaptation time decreases with increasing input steps due to the increased ligand gradient and flux.
On the contrary, after a negative step, the state of the receptor is {\it on} with a high intracellular concentration of ligand, and thus for the receptor to switch off, the intracellular concentration has to decrease below the (small) intracellular $K_{\it on}$. In this case, the larger the initial intracellular concentration the greater the time required to reach a small intracellular ligand concentration; the adaptation time consequently increases.

The experimental adaptation times behave very differently (see insets of Fig.~\ref{fig:times_FCD}a,b for a comparison with the slowest adaptation time course of the one-component model).
In particular, in most of the experimental data, the adaptation time in response to a positive step tends to increase with increasing step size (although activated RasG in {\it Dictyostelium} chemotaxis has the opposite trend \cite{Takeda_2012}).
In bacterial chemotaxis, this trend can be traced back to a maximal, saturated rate of receptor modification during adaptation \cite{Endres_2006_assistance_neighborhood,Keymer_2006}. Interestingly, the bacterial chemotaxis data we considered in response to a negative step exhibit a ``stereotypical response'', with an adaptation time independent of the amplitude of the stimulus, reflecting a more complicated and highly activated demethylation reaction \cite{Clausznitzer_2010, Min_2012}. Finally, Fig.~\ref{fig:times_FCD}c,d shows that fold-change detection is almost perfectly satisfied by our minimal model, demonstrating that even a simple model is capable of producing sophisticated sensory features.

\subsection*{Precision, sensitivity and response time}
\begin{figure}[htp]
 \begin{center}
\includegraphics{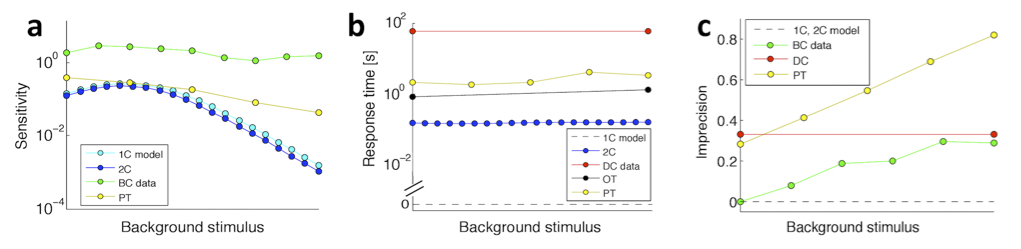}
\caption{{\bf Sensitivity, response time and imprecision.} (a) Sensitivity, as defined in Fig.~\ref{fig:pathways}, for the one-component (light blue) and two-component (dark blue) models in response to a step increase of 50\% in concentration, with prestimulus values from 0.01 to 100 mM. Bacterial chemotaxis (BC) and photo-transduction (PT) data of, respectively, receptor activity from {\it in vivo} FRET and transmembrane current are taken from Fig.~3B of \cite{Sourjik_2002} (20\% step change) and Fig. 6B of \cite{Forti_1989}. (b) Response time, as defined in Fig.~\ref{fig:pathways}, and its dependence on the background stimulus. Shown are the one-component model (dashed line), two-component model (dark blue), as well as {\it Dictyostelium} data (DC) of ACA calculated from Fig.~4C of \cite{Brzostowski_2013}, olfactory-transduction transmembrane current data (OT) from Fig~1a of \cite{Menini_1995}, and photo-transduction transmembrane current data (PT) from Fig. 6B of \cite{Forti_1989}.
(c) Imprecision, as defined in Fig.~\ref{fig:pathways}, for the one- and two-component model, BC data in response to MeAsp from Fig.~2c of \cite{Neumann_2010}, {\it Dictyostelium} data (DC) of ACA from Fig.~4C of \cite{Brzostowski_2013} and transmembrane current PT data from Fig. 6B of \cite{Forti_1989}.
}
 \label{fig:sens_rt}
 \end{center}
 \end{figure}

To further compare our minimal adaptation models with data, we consider three additional characteristics of adaptation. \textit{Sensitivity} represents the relative change of the output response with respect to a change in input stimulus (see Fig.~\ref{fig:pathways}). Both our one- and two-component models display small sensitivity values and dynamic ranges when compared with the experimental data available for bacterial chemotaxis and photo-transduction (see Fig.~\ref{fig:sens_rt}a).
This discrepancy can be understood considering that chemoreceptors in bacteria are known to cluster to increase their sensitivity. Additionally,  receptor types with different ligand-binding strengths are known to extend the dynamic range of sensing \cite{Endres_2006_assistance_neighborhood,Keymer_2006}.
These strategies could also be exploited in our equilibrium-physics models, but making clusters of different receptor types would nonetheless cost energy for the cell, even if this is paid for during cell growth and thus is not directly connected with sensing.

The \textit{response time} is the time interval between the onset of the stimulus step and the peak amplitude of the response (see Fig.~\ref{fig:pathways}). Figure~\ref{fig:sens_rt}b shows that the response of the one-component model is instantaneous, and is therefore similar to the fast bacterial chemotaxis response, while the two-component model resembles the slower eukaryotic responses.
Both in our one-component model and bacterial chemotaxis, the response is mainly determined by a ``conformational'' change in the receptor, which is very fast (ns-$\mu$s) \cite{Gegner_1992}. In contrast, eukaryotic responses involve long cascades based on diffusion.
Consistently, in both our simulations and the experimental data, the response time does not depend on the background stimulus, as this time is determined by the speed of cellular components.

Considering the steady state after adaptation to a step change, we can distinguish the level of precision of adaptation. In particular, if we define the \textit{imprecision} as presented in Fig.~\ref{fig:pathways}, we notice that both our one- and two-component models are perfectly precise, i.e. the steady state of the system is independent of the stimulus strength, even for large stimuli (Fig.~\ref{fig:sens_rt}c). This does not occur for bacterial chemotaxis, which is perfectly adapting for small background stimuli but loses precision with increasing stimulus strength.
While our minimal models are precisely adapted when fully equilibrated, precision in bacterial chemotaxis is regulated by a non-equilibrium pathway with constraints, e.g. from the finite number of methylation sites.
The same trend of imprecision is present in photo-transduction, which is not even fully precise at small background stimuli. This may be explained considering that the photo- and olfactory-transduction pathways (see Fig.~1 of \cite{Menini_1995}) represents only the first stage of a complex response, and consequently the output signal of these pathways undergoes further processing and error corrections. 
The {\it Dictyostelium} adenylyl cyclase activity shows a constant 33\% imprecision independent of stimulus strengths (Fig.~\ref{fig:sens_rt}c).
Note however that cGMP (Fig.~2A of \cite{Valkema_1994}) and activated RasG (Fig.~2A of \cite{Takeda_2012}) exhibit near perfect adaptation (data not shown in Fig.~\ref{fig:sens_rt}c).

\subsection*{Spatial gradient sensing}
\begin{figure}[htp]
 \begin{center}
\includegraphics{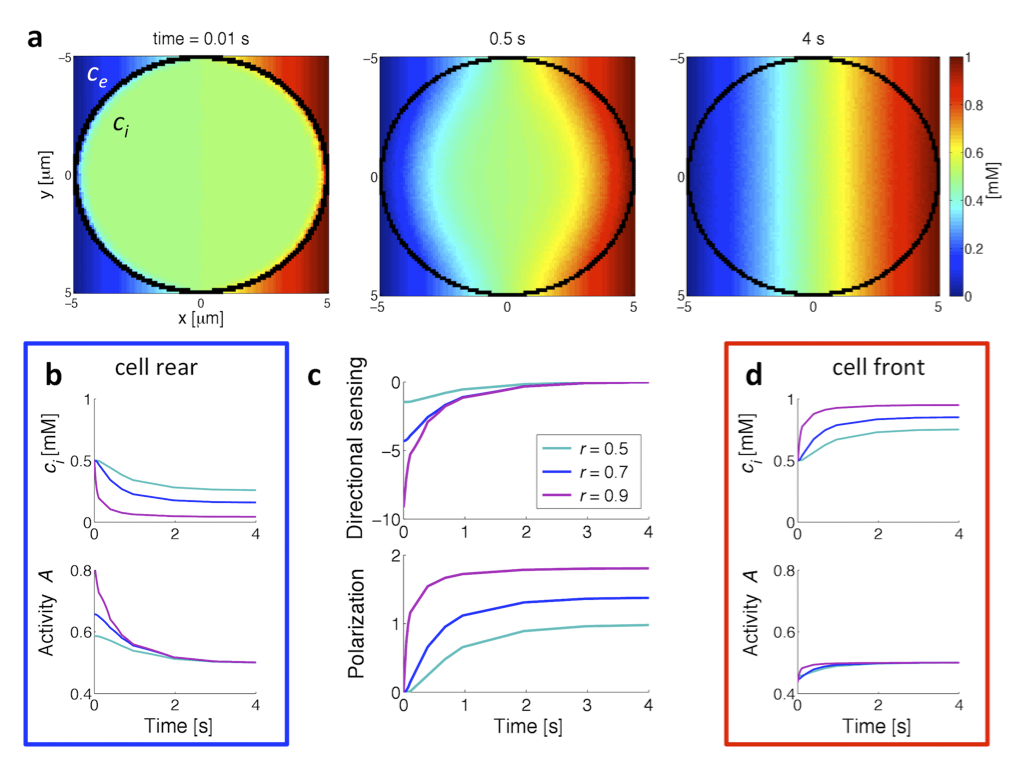}
\caption{{\bf Gradient sensing in the one-component model.} (a) Intracellular ($c_i$) and extracellular ($c_e$) ligand concentration in the $x$-$y$ plane for different times. Initially, the concentrations are homogeneously 0.5 mM. When the linear gradient is applied, this is changed to 0 mM at the cell rear at (-5 $\mu$m, 0) and 1 mM at the cell front at (5 $\mu$m, 0). (b) Internal concentration (top) and activity (bottom) time courses for different receptor lengths at the cell rear. (c) Directional sensing (top), as defined in Eq.~\eqref{dir_sens}, and polarization (bottom) for different receptor lengths. Polarization is defined as the difference between internal concentrations at positions (-5 $\mu m+l_{rec}$, 0) and (5 $\mu m-l_{rec}$, 0), normalized by the concentration at (0, 0). (d) Internal concentration and activity time courses for different receptor lengths at the cell front. See {\it Methods} for a more detailed description.
}
 \label{fig:gradient}
 \end{center}
 \end{figure}

When considering adaptation in {\it Dictyostelium} chemotaxis, it is worth noting that cells do not respond to step changes but to spatial gradients.
In particular, even if the models we are considering do not include cell motility, we can nevertheless study the response to those stimuli: approximating the cell by a round circle in a 2D plane with an initial homogeneous internal ligand concentration, we simulated the response of the one-component model when the external ligand concentration changes linearly in space across the cell length.
Figure~\ref{fig:gradient}a shows the spatial distribution of the attractant at different times due to slow diffusion across the membrane.
The internal concentration and receptor-activity time courses at the cell rear (minimal external concentration) and at the cell front (maximal external concentration) for different receptor lengths are depicted in Fig.~\ref{fig:gradient}b and d, respectively.
Also in spatial sensing the activity of the receptors adapts perfectly along the cell circumference.

To quantify directional sensing we consider the dipole moment $\mu$ of the receptor activity, defined as the sum of the activity on the cell circumference weighted by the normalized $x$ position along the gradient: 
\begin{equation}
\mu=\oint \frac{x}{r_0} [A(t)-A_{ss}],
\label{dir_sens}
\end{equation}
where $A_{ss}$ represents the adapted steady-state activity.
The initial response of $\mu$ is strong but then vanishes completely with adaptation of the receptors (see Fig.~\ref{fig:gradient}c, top).
Although the response ceases, the internal gradient remains, thus representing the cell's degree of polarization (Fig.~\ref{fig:gradient}c, bottom). The results of the two-component model are shown in Text S1.


\section*{Discussion}


In this work we analyzed and compared adaptation pathways from very different organisms, ranging from bacteria to eukaryotes. All these pathways require energy in order to adapt \cite{Lan_2012}.
Here, we showed that it is possible to build minimal adaptation mechanisms without the need of energy consumption by the cell, as possibly relevant for ancient protocells.
Despite their extreme simplicity, our two minimal models can help elucidate some aspects of complex signaling pathways.

Known transmembrane receptors in eukaryotes are grouped into ionotropic and metabotropic receptor types. Ionotropic receptors are characterized by a direct response, much as an ionic channel changing its conformation (such as opening or closing) in response to an extracellular stimulus. A direct activation of this kind is similar to the conformational change-based mechanism in the bacterial chemotaxis pathway.
In contrast, metabotropic receptors are more sophisticated: triggering of the receptor activates a cascade, usually a G-protein, leading to a change of second messenger concentration. This often involves complex feedback mechanisms \cite{Manahan_2004}. The influx/efflux of these second messengers, together with the current flowing through any ion channel activated by them, produce a change in membrane potential, which usually represents the output of the pathway.
All the eukaryotic examples presented in this work fall into this category (Fig.~\ref{fig:pathways}b-d).

Our two minimal models directly relate to metabo\-tropic signaling pathways.
The one-component model functions by means of a small diffusible ligand, which can permeate through the membrane.
This is somewhat comparable to the presence of cAMP both inside and outside starving {\it Dictyostelium} cells and to the inflow of ${\rm Ca^{2+}}$ in photo- and olfactory-transduction.
In addition, in both our one-component model and the {\it Dictyostelium} pathway this ligand is responsible for both sensing {\it and} adaptation.
cAMP is not only outside {\it and} inside the cell, but also follows the external stimulus (Fig.~5B of \cite{Brzostowski_2013}) and thus can be considered mediating adaptation of the adenylyl cyclase ACA (Fig.~4C in \cite{Brzostowski_2013}).
In contrast, the two-component model involves the detaching of the two subunits, $a$ and $b$, from the receptor after its activation, precisely resembling the dissociation of the $\alpha$ and the $\beta\gamma$ subunits in {\it Dictyostelium}, photo- and olfactory-transduction pathways.
Similar to the cAR1 receptor and G-protein in {\it Dictyostelium} \cite{Janetopoulos_2001}, the first component of the two-component model does not adapt, leaving the adapting response to the second component, the latter resembling the time course of adenylyl cyclase ACA.
The network motif effectively implemented in both minimal models is the incoherent feedforward loop (see Fig.~\ref{fig:models}, right panels), which is encountered in the {\it Dictyostelium} pathway as well \cite{Takeda_2012}. This design principle can easily be identified through the presence of a slow inhibitory process, which is the transmembrane diffusion of the ligand in the one-component model and the diffusion of the $b$ subunit in the two-component model.

An important difference between the biological signaling pathways and our minimal models is the source of energy dissipation.
Unlike our minimal models, cells have to pay for their significant energy costs. Hence, what are the advantages which may have led to the evolution of biological non-equilibrium pathways?
A first drawback of our models is that they exhibit a low sensitivity and dynamic range (Fig.~\ref{fig:sens_rt}a).
However, as already mentioned, this could be amended by introducing receptor complexes of different receptor types \cite{Keymer_2006}.
A more serious constraint of our equilibrium models is that the response and adaptation times are determined by diffusion constants which cannot easily be adjusted by the cell.
Furthermore, the one-component model requires the external stimulus to enter the cell, while modern energy-consuming pathways generally separate external sensing from internal signaling, thus avoiding that toxic chemicals enter the cell to mediate adaptation.

Why do G-protein cascades employ small fast diffusible mole\-cules with little spatial control to mediate adaptation? A possibility is stimulus amplification since active G-protein subunits can further activate many downstream signaling molecules \cite{Ramanathan_2005,vanHemert_2010}. In addition, eukaryotic cells are often highly specialized, as in the case of olfactory receptor neurons and photoreceptors, and thus the low specificity of these small molecules is compensated by the high specificity of the cell types. Alterations in transmembrane potential also permits fast and reliable electrical transmission through excitation, typical of neurons.

Some molecular species of our minimal models may represent ``fossils'', remnants of ancient protocells in current adaptation pathways.
For instance, the role of the ``non activating'' G-protein subunit remains unclear in eukaryotic signaling \cite{McMains_2008}. 
Others may have taken on new roles: GTP binding and hydrolysis may have introduced a ``timer'' into the pathways, promoting the reassociation of the G-protein complex and thus the termination of the downstream activation.
The consequence may be a bursty, frequency modulated signaling, with the advantage of being more accurate for both sensing and encoding \cite{Tostevin_2012,Mora_2010,Tu_2008}.

In conclusion, our simple schemes for perfect adaptation are energy efficient, but evolution may have replaced them by energy-consuming pathways to increase adjustability and control of the response and adaptation times for the cells' changing needs.
Similar to kinetic proofreading, in which the probability of a correct output is increased through repeated cycles \cite{Hopfield_1974,Murugan_2012,Franccois_2013}, adaptation pathways could represent schemes in which cells improve the control and robustness of the response by exploiting energy expenditure for enhanced fitness.


\section*{Methods}

{\bf Energy dissipation of one-component model.}
For a process described by forward ($r_+$) and reverse ($r_-$) rates, the entropy production rate is given by \cite{Qian_2007}
\begin{equation}
\frac{dS}{dt}=(r_+-r_-)\ln\left(\frac{r_+}{r_-}\right)\geqslant 0, 
\label{eq:entropy_general}
\end{equation}
with entropy in units of the Boltzmann constant $k_B$.
For our one-component model, following Fick's first law, $r_+=D c_e(t)/h_m$ and $r_-=D c_i(t)/h_m$, with $D= 3$ $\mu$m$^2$/s the diffusion constant of the ligand, and $h_m=10$ nm the membrane thickness, leading for the total cell of radius $r_0$ to
\begin{equation}
 \frac{dS}{dt}=4\pi r_0^2 D  \left[\frac{c_e(t)-c_i(t)}{h_m}\right]\ln\left(\frac{c_e(t)}{c_i(t)}\right),
\label{eq:entropy_models}
\end{equation}
which is equal to 0 at steady state (i.e. when $c_e=c_i$), and $> 0$ only when a concentration gradient across the membrane is present. The energy dissipation rate corresponds to Eq.~\eqref{eq:entropy_models} multiplied by the temperature $T$ of the system.

{\bf Energy dissipation vs precision in bacterial chemotaxis.}
Equation~\eqref{eq:chem_model} for precise adaptation \cite{Clausznitzer_2010} can be generalized to include both the forward and reverse reactions
\begin{eqnarray}
&\left.\frac{dm}{dt}\right|_{\rm nonequ}=&g_R(1-A)-g_{-R}A-[g_B A-g_{-B}(1-A)], 
\label{eq:bact_precise}
\end{eqnarray}
where the first two terms represent the contribution of CheR, and the last two the contribution of CheB. The reverse rate constants, $g_{-R}$ and $g_{-B}$, can be adjusted to keep the net fluxes $g_R(1-A)-g_{-R}A \gg 0$ and $g_B A-g_{-B}(1-A) \gg 0$ at steady state, such that the net actions of CheR and CheB are methylation and demethylation, respectively. We used parameters $g_R=0.0069$ ${\rm s^{-1}}$ \cite{Clausznitzer_2010}, $g_{-R}=0.1$ $g_R$, $g_B=0.11$ ${\rm s^{-1}}$ \cite{Clausznitzer_2010}, $g_{-B}=0.1$ $g_B$. 
To describe imprecise adaptation in response to a stimulus of concentration $c$, we consider
\begin{eqnarray}
&\left.\frac{dm}{dt}\right|_{\rm equ}=&l_R c(m_{\rm max}-m)-l_{-R} c m-[l_B m-l_{-B}(m_{\rm max}-m)],
\label{eq:bact_imprecise}
\end{eqnarray}
which does not dissipate energy when rate constants $l_R$, $l_{-R}$, $l_B$, $l_{-B}$ fulfill equilibration conditions $l_R (m_{max}-m)-l_{-R} m=0$ and $l_B m-l_{-B}(m_{max}-m)=0$. Here, we chose $l_R=l_{-R}=0.1$ ${\rm M^{-1}s^{-1}}$ and $l_B=l_{-B}=0.001$ ${\rm s^{-1}}$ (this leads to equilibrium for $m=m_{max}/2$). Since the dynamics of the methylation level does not depend on the activity, Eq.~\eqref{eq:bact_imprecise} leads to imprecise adaptation.
Equations~\eqref{eq:bact_precise} and \eqref{eq:bact_imprecise} can be considered two components of the same system. By combining them through a parameter $w$, we can describe their relative contributions
\begin{equation}
\frac{dm}{dt}=(1-w)\left.\frac{dm}{dt}\right|_{\rm nonequ}+w\left.\frac{dm}{dt}\right|_{\rm equ}.
\label{eq:bact_total}
\end{equation}
Following Eq.~\eqref{eq:entropy_general}, summing over the different reactions and neglecting the contributions of phosphorylation of CheB and CheY, the corresponding entropy-production rate is given by
\begin{equation}
\frac{dS}{dt}=N_{\rm rec}\left[(1-w)\left.\frac{dS}{dt}\right|_{\rm nonequ}+w\left.\frac{dS}{dt}\right|_{\rm equ}\right],
\label{eq:entropy_total}
\end{equation}
with $N_{\rm rec}=15,000$ the approximate number of receptors in a bacterium \cite{Li_2004}. By moving $w$ from 0 to 1, this system becomes gradually imprecise and approaches equilibrium (see Fig.~\ref{fig:entropy}b).

{\bf Gradient sensing simulations.}
To simulate the response of the one- and two-component models we considered a two-dimensional cell and numerically solved the diffusion equation where the boundary conditions at a distance $r_0$ from the center are given by a gradient in the $x$ direction, with 0 mM corresponding to the minimal concentration at $x=-5$ $\mu$m and 1 mM corresponding to the maximal concentration at $x=5$ $\mu$m. Both the creation of the mesh and the solution of the equation were obtained by means of the Partial Differential Equation Toolbox of MATLAB (The MathWorks, Inc., Natick, Massachusetts, United States).

\section*{Acknowledgments}
G.D.P. and R.G.E. were supported by ERC Starting Grant 280492-PPHPI.

%

%
%

\newpage


%
%

\end{document}